\documentclass[twocolumn,dvipdfmx,showpacs,preprintnumbers,amsmath,amssymb,aps,prd,nofootinbib,superscriptaddress,eqsecnum]{revtex4}
\usepackage{makeidx}
\usepackage[dvipdfmx]{graphicx}
\usepackage[all]{xy}
\usepackage{feynmp}
\usepackage{color}
\usepackage{amsmath,amssymb,bm}

\newcommand{\todaye}{\the\year /\the\month /\the\day}

\newcommand{\mc}{\mathcal}
\newcommand{\bee}[1]{\begin{align} #1 \end{align}}
\newcommand{\mt}[1]{\begin{matrix} #1 \end{matrix}}
\newcommand{\nn}{\nonumber}
\newcommand{\lrax}[2]{\overset{#1}{\underset{#2}{\leftrightarrows}}}
\newcommand{\tr}{\mathrm{Tr}}
\newcommand{\Dsl}{D \hspace{-2.2mm}/}
\newcommand{\braa}[1]{\left( #1 \right)}
\newcommand{\brab}[1]{\left\{ #1 \right\}}
\newcommand{\brac}[1]{\left[ #1 \right]}
\newcommand{\brad}[1]{\left| #1 \right|}
\newcommand{\brae}[1]{\left\langle #1 \right\rangle}

\begin{document}

\title{
Extended Goldberger-Treiman relation obtained in a three-flavor parity doublet model
}
\author{Hiroki Nishihara\footnote{h248ra@hken.phys.nagoya-u.ac.jp}}
\author{Masayasu Harada\footnote{harada@hken.phys.nagoya-u.ac.jp}}
\affiliation{
Department of Physics, Nagoya University, Nagoya 464-8602, Japan
}

\date{\today}

\def\theequation{\thesection.\arabic{equation}}
\makeatother

 \begin{abstract}
We study masses and decay widths of positive and negative parity nucleons using a three-flavor parity doublet model, in which we introduce three representations,
$\left[({\bf 3} , \bar{{\bf 3}})\oplus (\bar{{\bf 3}} , {\bf 3})\right]$, $\left[({\bf 3} , {\bf 6}) \oplus ({\bf 6}, {\bf 3})\right]$, and $\left[({\bf 8} , {\bf 1}) \oplus ({\bf 1} , {\bf 8})\right]$ 
of the chiral U$(3)_{\rm L}\times$U$(3)_{\rm R}$ symmetry.
We find an extended version of the Goldberger-Treiman relation among the mass differences and the coupling constants for pionic transitions.
This relation leads to
an upper bound for the decay width of $N(1440) \rightarrow N(939) + \pi$ independently of the model parameters.
We perform the numerical fitting of
the model parameters and 
derive several predictions, 
which can be tested in future experiments or lattice QCD analysis.
Furthermore, when we use the axial coupling of the excited nucleons obtained from lattice QCD analyses as inputs,
we find that the ground state nucleon $N(939)$ consists of about 80\% of $\left[({\bf 3} , {\bf 6}) \oplus ({\bf 6}, {\bf 3})\right]$ 
component, and that 
the chiral invariant mass of $N(939)$ is roughly $500$ -- $800$\,MeV.
\end{abstract}

\pacs{
14.20.Gk,
12.39.Fe,
13.30.Eg,
14.20.Dh
}

\maketitle

\section{Introduction}
\label{sec:Introduction}

There are many baryons in Nature~\cite{Agashe:2014kda}, which are described by
the flavor symmetry as in Table~\ref{quark model1}. 
It is considered that the Quantum ChromoDynamics (QCD) describes their properties such as the masses and the interactions as well as the structure under the flavor symmetry.
One of the most important features of QCD relevant to the low-energy hadron physics is the chiral symmetry and its spontaneous breaking.
The spontaneous chiral symmetry breaking generates mass differences between the chiral partners, which lead to the mass difference between the parity partners, as well as the mixing among different chiral representations.
It is interesting 
to study the role of the chiral symmetry breaking to determine the properties and structures of baryons.
In particular,  asking how much amount of the masses of baryons are generated by the chiral symmetry breaking is an attractive question.
\begin{table}
\begin{center}
\caption[]{Data of spin 1/2 baryons in PDG~\cite{Agashe:2014kda}.}
\begin{tabular}{c|cccc}\hline\hline
\hspace{2mm}$J^P$\hspace{2mm}
&&Octets& &
\\\hline
${1/2}^+$
&
$N(939)$
&
$ \Lambda(1116) $
&
$\Sigma(1193)$
&
$ \Xi(1318)$
\\
${1/2}^+$
&
$N(1440) $
&
$\Lambda(1600)$
&
$ \Sigma(1660) $
&
$\Xi(1690)$
\\
${1/2}^-$
&
$N(1535) $
&
$\Lambda(1670) $
&
$\Sigma(1560) $
&
$\Xi(?)$
\\
${1/2}^-$
&
$N(1650) $
&
$\Lambda(1800) $
&
$\Sigma(1620) $
&
$\Xi(?)$
\\
${1/2}^+$
&
$N(1710) $
&
$\Lambda(1810) $
&
$\Sigma(1880) $
&
$\Xi(?)$
\\\hline\hline
\end{tabular}
\end{center}
\label{quark model1}
\end{table}

Two-flavor parity doublet model was proposed by Ref.~\cite{Detar:1988kn} in which the nucleon $N(939)$ $\left(J^P=\frac{1}{2}^{+}\right)$ and its parity partner $N(1535)$ $\left(J^P=\frac{1}{2}^{-}\right)$ are introduced as a doublet.
This model includes a chiral invariant mass denoted by $m_0$, which implies that 
the masses of $N(939)$ and $N(1535)$ tend to $m_0$ when the chiral symmetry is restored and 
their mass splitting is given by the spontaneous chiral symmetry braking.
This structure has been studied by many works~\cite{Detar:1988kn,Nemoto:1998um,
Jido:1998av,Jido:1999hd,Jido:2001nt,Chen:2009sf,Chen:2010ba,Chen:2011rh,Glozman:2012fj,Aarts:2015mma,Gallas:2013ipa}.
One of the key feature of the structure is a relation among the pionic interactions and mass difference between the chiral partners, which is an extended Goldberger-Treiman relation.
Using the relation, one can obtain some information of $m_0$ by comparing the model with experimental data, which seems to prefer small value of $m_0$, $m_0 \lesssim 500$\,MeV~(See e.g. Refs.~\cite{Jido:2001nt,Gallas:2013ipa}).
On the other hand, 
parity doublet models are applied to study the nuclear matter, where large value of $m_0$ (e.g. $m_0 \gtrsim 800$\,MeV~\cite{Zschiesche:2006zj,Dexheimer:2007tn}, $m_0 \gtrsim 500$\,MeV~\cite{Motohiro:2015taa}) seems to be preferred.
The results by the lattice QCD analysis in Refs.~\cite{Glozman:2012fj,Aarts:2015mma} indicate that most part of the mass of the nucleon is the chiral invariant mass $m_0$.

References~\cite{Chen:2009sf,Chen:2010ba,Chen:2011rh} 
gave the extension to the three-flavor parity doublet model 
 with the chiral 
U$(3)_{\rm L}$$\times$U$(3)_{\rm R}$
symmetry.
In two-flavor case, the axial charge $g_A^{}$ of the nucleon is constrained as $g_A \leq 1$ as mentioned in Ref.~\cite{Detar:1988kn}. However, as is well known, it is empirically determined as $g_A=F+D=1.268\pm 0.006$,  where $F$ and $D$ are the axial coupling constants given in Ref.~\cite{Cabibbo:1963yz}, and their values are $ 0.475 \pm 0.004$ and $0.793 \pm 0.005$ \cite{Yamanishi:2007zza}. 
To solve this puzzle, the authors in Refs.~\cite{Chen:2009sf,Chen:2010ba,Chen:2011rh}  introduced three parity-even and three parity-odd baryons which belong to $\brac{({\bf 3} , \bar{{\bf 3}})\oplus (\bar{{\bf 3}} , {\bf 3})}$, $\brac{({\bf 3} , {\bf 6}) \oplus ({\bf 6}, {\bf 3})}$, and $\brac{({\bf 8} , {\bf 1}) \oplus ({\bf 1} , {\bf 8})}$ representations of the chiral U$(3)_{\rm L}\times$U$(3)_{\rm R}$ group, and showed that the baryon mass spectra and the axial coupling constants are reasonably agree with experiments when the number representations are suitably restricted.

In this paper, we make a general analysis using a three-flavor parity doublet model including all possible representations to fit available experimental data.
We will show that there exists an extended Goldberger-Treiman relation among the masses and pionic coupling constants for the relevant nucleons, from which we obtain an upper bound for the $N(1440) \to N(939) + \pi$ decay width independently of the values of model parameters.
Furthermore, our numerical analysis shows that the ground state nucleon $N(939)$ includes about 80\% of $\brac{({\bf 3} , {\bf 6}) \oplus ({\bf 6}, {\bf 3})}$ 
component, and that the chiral invariant mass of $N(939)$ is roughly $500$  -- $800$\,MeV consistently with the analysis~\cite{Motohiro:2015taa}.

This paper is organized by five sections: 
In Sec.~\ref{sec:model} we construct a three-flavor parity doublet model, and give the mass matrix for the nucleons.
In Sec.~\ref{sec:upper bounds} we show an extended Goldberger-Treiman relation and obtain the upper bound of $\Gamma (N(1440) \rightarrow N(939) + \pi)$.
Section \ref{sec:Numerical fitting} gives the numerical fitting results for the masses and the partial decay widths of the nucleon and the exited nucleons. 
Finally, we will give a brief summary and discussions in Sec.~\ref{sec:summary}.

\section{Three-flavor parity doublet model}
\label{sec:model}
In this section we introduce the baryon fields with parity doublet structure and construct a model Lagrangian for the baryons interacting with the scalar and pseudoscalar mesons,
requiring the U$(3)_{\rm L}$$\times$U$(3)_{\rm R}$ chiral invariance.
Although the notations used here are somewhat different, the construction itself is consistent with Ref.~\cite{Chen:2009sf}. 

\subsection{Construction of the Lagrangian}

The chiral representations of quarks under $\mbox{SU(3)}_{\rm L} \times \mbox{SU(3)}_{\rm R}$ are written as
\bee{
q_L \sim ({\bf 3}, {\bf 1})
\ ,~~
q_R \sim ({\bf 1}, {\bf 3}) \ ,
}
where these ${\bf 3}$ and ${\bf 1}$ in the bracket express triplet and singlet for SU$(3)$, respectively. They transform under the U$(1)_{\rm L}$$\times$U$(1)_{\rm R}$ symmetry as
\bee{
q_L &\rightarrow e^{i\theta_L}q_L\equiv e^{i\frac{1}{3}\theta_V -i\theta_A }q_L
\ ,
\nn\\
q_R &\rightarrow e^{i\theta_R}q_R\equiv e^{i\frac{1}{3}\theta_V +i\theta_A }q_R
\ ,}
where $\theta_V$ and $\theta_A$ are the transformation parameters.
Here we assign the U$(1)_{\rm A}$ charge of the left- and right-handed quarks  as $-1$ and $+1$, respectively.
In the following we explicitly write the $\mbox{U(1)}_{\rm A}$ charge as
\begin{equation}
q_L \sim ({\bf 3}, {\bf 1})_{-1}
\ ,~~
q_R \sim ({\bf 1}, {\bf 3})_{+1}\ .
\end{equation}
Since a baryon can be expressed as a direct product of three quarks, we have the following possibilities for the chiral representations of baryons:
\begin{align}
q_L \otimes q_R \otimes q_R
&\sim ({\bf 3}, {\bf 1})_{-1} \otimes ({\bf 1}, {\bf 3})_{+1} \otimes ({\bf 1}, {\bf 3})_{+1}
\notag\\
&\sim ({\bf 3}, \bar{{\bf 3}})_{+1} \oplus ({\bf 3}, {\bf 6})_{+1}
\ ,
\\
q_L \otimes q_L \otimes q_L
&\sim ({\bf 3}, {\bf 1})_{-1} \otimes ({\bf 3}, {\bf 1})_{-1} \otimes ({\bf 3}, {\bf 1})_{-1}
\notag\\
&\sim  ({\bf 1}, {\bf 1})_{-3} \oplus 2({\bf 8}, {\bf 1})_{-3}  \oplus ({\bf 10}, {\bf 1})_{-3}
\ ,
\end{align}
and ones with $q_L$ and $q_R$ replaced with each other.
After the chiral symmetry is spontaneously broken down to the flavor $\mbox{SU(3)}_{\rm F}$ symmetry, octet-baryons appear from the representations of $ ({\bf 3}, \bar{{\bf 3}})_{-1}$, $({\bf 3} , {\bf 6})_{-1},$ and $({\bf 8}, {\bf 1})_{-3}$.
Therefore we introduce three types of baryons corresponding these three representations.
Hereafter, for simplify, we focus only on baryon which has $1/2$ spin.

We introduce the following baryons:
\bee{
&
\Psi_{1l} 
\sim ( {\bf 3}, \bar{{\bf 3}})_{+1} 
\ ,~~
\Psi_{1r} 
\sim (\bar{{\bf 3}}, {\bf 3})_{-1} 
\ ,~~
\nn\\&
\Psi_{2l} 
\sim (\bar{{\bf 3}}, {\bf 3})_{-1} 
\ ,~~
\Psi_{2r} 
\sim ( {\bf 3}, \bar{{\bf 3}})_{+1}  
\ ,
\nn\\&
\eta_{1l}\, 
\sim ({\bf 3}, {\bf 6})_{+1} 
\ ,~~
\eta_{1r}\ 
\sim ({\bf 6}, {\bf 3})_{-1} 
\ ,~~
\nn\\&
\eta_{2l}\, 
\sim ({\bf 6}, {\bf 3})_{-1} 
\ ,~~
\eta_{2r}\ 
\sim ({\bf 3}, {\bf 6})_{+1} 
\ ,
\nn\\&
\chi_{1l} 
\sim \braa{{\bf 8} , {\bf 1}}_{-3} 
\ ,~~
\chi_{1r} 
\sim \braa{{\bf 1}, {\bf 8}}_{+3}
\ ,~~
\nn\\&
\chi_{2l} 
\sim \braa{{\bf 1}, {\bf 8}}_{+3}
\ ,~~
\chi_{2r} 
\sim \braa{{\bf 8} , {\bf 1}}_{-3}
\ ,
\label{reps}
}
which are listed in Table~\ref{low-chiral-trans} together with $M$ defined later.
\begin{table*}[htb]
\caption[]{Chiral representation.
}
\vspace{-3mm}
\begin{center}
\begin{tabular}{c|ccccccccccccc}\hline\hline
fields&
$\Psi_{1l}$&$\Psi_{1r}$&
$\Psi_{2l}$&$\Psi_{2r}$&
$\eta_{1l}$&$\eta_{1r}$&
$\eta_{2l}$&$\eta_{2r}$&
$\chi_{1l}$&$\chi_{1r}$&
$\chi_{2l}$&$\chi_{2r}$&
$M$
\\\hline
SU$(3)_{\rm L}$$\times$SU$(3)_{\rm R}$
&
$\braa{ {\bf 3} ,\bar{{\bf 3}}}$&
$\braa{ \bar{{\bf 3}} , {\bf 3}}$&
$\braa{ \bar{{\bf 3}} , {\bf 3}}$&
$\braa{ {\bf 3}, \bar{{\bf 3}} }$&
$\braa{ {\bf 3} ,{\bf 6}}$&
$\braa{ {\bf 6} , {\bf 3}}$&
$\braa{ {\bf 6} , {\bf 3}}$&
$\braa{ {\bf 3} ,{\bf 6}}$&
$\braa{ {\bf 8} , {\bf 1}}$&
$\braa{ {\bf 1} , {\bf 8}}$&
$\braa{ {\bf 1} , {\bf 8}}$&
$\braa{ {\bf 8} , {\bf 1}}$&
$\braa{ {\bf 3} ,\bar{{\bf 3}}}$
\\
U$(1)_{\rm A}$ charge&
$+1$&$-1$&
$-1$&$+1$&
$+1$&$-1$&
$-1$&$+1$&
$-3$&$+3$&
$+3$&$-3$&
$-2$
\\\hline\hline
\end{tabular}
\end{center}
\label{low-chiral-trans}
\caption[]{Allowed operators with one meson field $M$ which has $-2$ as the U$(1)_{\rm A}$ charge. ``$\times$'' means that such operators are forbidden. }
\vspace{-3mm}
\begin{center}
\begin{tabular}{cc|cccc|cccc|cccc}\hline\hline
fields&-&
$\Psi_{1l}$&$\Psi_{1r}$&$\Psi_{2l}$&$\Psi_{2r}$&$\eta_{1l}$&$\eta_{1r}$&$\eta_{2l}$&$\eta_{2r}$&
$\chi_{1l}$&$\chi_{1r}$&$\chi_{2l}$&$\chi_{2r}$
\\
-&U$(1)_{\rm A}$&${+1}$&${-1}$&${-1}$&${+1}$&${+1}$&${-1}$&${-1}$&${+1}$&${-3}$&${+3}$&${+3}$&${-3}$
\\\hline
$\bar{\Psi}_{1l}$&${-1}$&
$\times$&$\mc{W}^{(1)}_{R}$&$\times$&$\times$&
$\times$&$\mc{O}^{(1)}_{R}$&$\times$&$\times$&
$\times$&$\mc{O}^{(2)}_{R}$&$\times$&$\times$\\
$\bar{\Psi}_{1r}$&${+1}$&
${\mc{W}}^{(1)}_L$&$\times$&$\times$&$\times$&
$\mc{O}^{(1)}_L$&$\times$&$\times$&$\times$&
$\mc{O}^{(2)}_L$&$\times$&$\times$&$\times$\\
$\bar{\Psi}_{2l}$&${+1}$&
$\times$&$\times$&$\times$&$\mc{W}^{(2)}_R$&
$\times$&$\times$&$\times$&$\mc{O}^{(3)}_R$&
$\times$&$\times$&$\times$&$\mc{O}^{(4)}_{R}$\\
$\bar{\Psi}_{2r}$&${-1}$&
$\times$&$\times$&${\mc{W}}^{(2)}_L$&$\times$&
$\times$&$\times$&$\mc{O}^{(3)}_L$&$\times$&
$\times$&$\times$&$\mc{O}^{(4)}_L$&$\times$\\\hline
$\bar{\eta}_{1l}$&${-1}$&
$\times$&${\mc{O}^{(1)}_L}^\dagger$&$\times$&$\times$&
$\times$&$\mc{W}^{(3)}_R$&$\times$&$\times$&
$\times$&$\mc{O}^{(5)}_R$&$\times$&$\times$
\\
$\bar{\eta}_{1r}$&${+1}$&
${\mc{O}^{(1)}_R}^\dagger$&$\times$&$\times$&$\times$&
$\mc{W}^{(3)}_L$&$\times$&$\times$&$\times$&
$\mc{O}^{(5)}_L$&$\times$&$\times$&$\times$
\\
$\bar{\eta}_{2l}$&${+1}$&
$\times$&$\times$&$\times$&${\mc{O}^{(3)}_L}^\dagger$&
$\times$&$\times$&$\times$&$\mc{W}^{(4)}_R$&
$\times$&$\times$&$\times$&$\mc{O}^{(6)}_R$
\\
$\bar{\eta}_{2r}$&${-1}$&
$\times$&$\times$&${\mc{O}^{(3)}_R}^\dagger$&$\times$&
$\times$&$\times$&$\mc{W}^{(4)}_L$&$\times$&
$\times$&$\times$&$\mc{O}^{(6)}_L$&$\times$
\\\hline
$\bar{\chi}_{1l}$&${+3}$&
$\times$&${\mc{O}^{(2)}_L}^\dagger$&$\times$&$\times$&
$\times$&${\mc{O}^{(5)}_L}^\dagger$&$\times$&$\times$&
$\times$&$\times$&$\times$&$\times$\\
$\bar{\chi}_{1r}$&${-3}$&
${\mc{O}^{(2)}_R}^\dagger$&$\times$&$\times$&$\times$&
${\mc{O}^{(5)}_R}^\dagger$&$\times$&$\times$&$\times$&
$\times$&$\times$&$\times$&$\times$\\
$\bar{\chi}_{2l}$&${-3}$&
$\times$&$\times$&$\times$&${\mc{O}^{(4)}_L}^\dagger$&
$\times$&$\times$&$\times$&${\mc{O}^{(6)}_L}^\dagger$&
$\times$&$\times$&$\times$&$\times$\\
$\bar{\chi}_{2r}$&${+3}$&
$\times$&$\times$&${\mc{O}^{(4)}_R}^\dagger$&$\times$&
$\times$&$\times$&${\mc{O}^{(6)}_R}^\dagger$&$\times$&
$\times$&$\times$&$\times$&$\times$
\\\hline\hline
\end{tabular}
\end{center}
\label{table:allowed operators}
\end{table*}
Their subscripts such as $l$ or $r$ express its chirality:
\bee{
\gamma_5\Psi_{il}=& -\Psi_{il}
\ ,~~
\gamma_5\Psi_{ir}= +\Psi_{ir}
\ ,
\nn\\
\gamma_5\eta_{il}=& -\eta_{il}
~\ ,~~
\gamma_5\eta_{ir}= +\eta_{ir}
\ ,
\nn\\
\gamma_5\chi_{il}=& -\chi_{il}
\ ,~~
\gamma_5\chi_{ir}= +\chi_{ir}
}
for $ i=1, 2 $.

For clarifying the representations under the chiral group,
we
explicitly write the superscripts of the baryon fields as
\begin{align}
&
\braa{\Psi_{1l} }^a_\alpha
\ ,~~
\braa{\Psi_{1r}}_a^\alpha
\ ,~~
\braa{\Psi_{2l}}_a^\alpha 
\ ,~~
\braa{\Psi_{2r}}^a_\alpha 
\ ,
\notag\\
&
\eta_{1l}^{(a, \alpha \beta)}\ 
\ ,~~
\eta_{1r}^{(ab, \alpha)}\ 
\ ,~~
\eta_{2l}^{(ab, \alpha)}\ 
\ ,~~
\eta_{2r}^{(a, \alpha \beta)}\ 
\ ,\notag\\
&
\braa{\chi_{1l}}^a_b ~
\ ,~~
\braa{\chi_{1r}}^\alpha_\beta 
\ ,~~
\braa{\chi_{2l}}^\alpha_\beta ~
\ ,~~
\braa{\chi_{2r}}^a_b 
\ ,
\end{align}
where $a,b = 1,2,3$ are for $\mbox{SU}(3)_{\rm L}$ and 
$\alpha,\beta=1,2,3$ for $\mbox{SU(3)}_{\rm R}$.
Note that the superscripts of $ab$ and $\alpha\beta$ are symmetrized to express 
$\boldsymbol{6}$ representation.

The property for the party transformation $(\mc{P})$ and the charge conjugation $(\mc{C})$ are defined as
\bee{
&
\psi_{1l, 1r} \xrightarrow[]{\mc{P}} \gamma_0~\psi_{1r, 1l}
\ ,~~~~~~
\psi_{2l, 2r} \xrightarrow[]{\mc{P}} -\gamma_0~\psi_{2r, 2l}
\ ,
\nn\\&
\psi_{1l, 1r} \xrightarrow[]{\mc{C}} C \braa{\bar{\psi}_{1r, 1l}}^T
\ ,~~~
\psi_{2l, 2r} \xrightarrow[]{\mc{C}} - C\braa{\bar{\psi}_{2r, 2l}}^T
}
for $\psi = \Psi,~ \eta,~\chi$ with $C\equiv i\gamma^2\gamma^0$.
This assignment for the parity is called as the mirror assignment.

Now the covariant derivatives for each field are expressed as
\bee{
D_\mu \Psi_{1l,2r}
&= \partial_\mu \Psi_{1l, 2r}  
- i\mc{L}_\mu  \Psi_{1l,2r}
+ i \Psi_{1l,2r} \mc{R}_\mu
\ ,\nn\\
D_\mu \Psi_{1r,2l}
&= \partial_\mu \Psi_{1r, 2l}  
- i\mc{R}_\mu  \Psi_{1r,2l}
+ i \Psi_{1r,2l} \mc{L}_\mu
\ ,\nn\\
\braa{D_\mu \eta_{1l, 2r}}^{(a, \alpha\beta)} 
&= \partial_\mu \eta_{1l, 2r}^{(a, \alpha\beta)}-i  \braa{ \mc{L}_\mu}^a_b  \eta_{1l, 2r}^{(b, \alpha\beta)}
\nn\\&~~~
- i\brac{ \braa{\mc{R}_\mu}^\alpha_\rho \delta^{\beta}_\sigma
+ \delta^\alpha_\rho\braa{\mc{R}_\mu}^{\beta}_\sigma    
}\eta_{1l,2r}^{(a, \rho \sigma)}
\ ,\nn\\
\braa{D_\mu \eta_{1r, 2l}}^{(ab, \alpha)} 
&= \partial_\mu \eta_{1r, 2l}^{(ab, \alpha)}
-i  \braa{ \mc{R}_\mu}^\alpha_\beta  \eta_{1r, 2l}^{(ab, \beta)}
\nn\\&~~~
- i\brac{ \braa{\mc{L}_\mu}^a_c \delta^b_d
+ \delta^a_c\braa{\mc{L}_\mu}^b_d
}\eta_{1r,2l}^{(cd, \alpha)}
\ ,\nn\\
D_\mu \chi_{1l,2r} 
&=
\partial_\mu \chi_{1l,2r}  
- i\mc{L}_\mu \chi_{1l,2r}
+ i\chi_{1l,2r} \mc{L}_\mu 
\ ,\nn\\
D_\mu \chi_{1r,2l} 
&=
\partial_\mu \chi_{1r,2l}  
- i\mc{R}_\mu \chi_{1r,2l}
+ i\chi_{1r,2l} \mc{R}_\mu 
} 
where
$ \mc{L}_\mu$ and $\mc{R}_\mu$ are 
the external gauge fields introduced by gaging the chiral SU$(3)_{\rm L}$$\times$SU$(3)_{\rm R}$
symmetry.

Next we introduce a $3 \times 3$ matrix field $M$ expressing a nonet of scalar and pseudoscalar mesons made of a quark and an anti-quark.
The representation under $\mbox{SU(3)}_{\rm L} \times \mbox{SU(3)}_{\rm R} \times \mbox{U(1)}_{\rm A}$ of the $M$ is 
\begin{equation}
M \sim (\boldsymbol{3} ,\, \bar{\boldsymbol{3}} )_{-2} \ ,
\end{equation}
which implies that this field is actually made from $\braa{\bar{q}_Rq_L}$.
The parity transformation and the charge conjugation are also defined as
\bee{
M \xrightarrow[]{\mc{P}} {M}^\dagger
\ ,~~
M \xrightarrow[]{\mc{C}} {M}^T
\ .
}

We construct a general Lagrangian invariant under the chiral U$(3)_{\rm L}$$\times$U$(3)_{\rm R}$  symmetry.~\footnote{%
$\mbox{U}(1)_{\rm A}$ symmetry is explicitly broken by the anomaly. As we show later, we  introduce a term proportional to $\ln \left( \det M / \det M^\dag \right)$ which saturates the anomaly, and assume that all the other terms of the effective model are invariant under the  $\mbox{U}(1)_{\rm A}$ transformation.
}
In the present analysis, we include only non-derivative Yukawa interactions among one meson field $M$ and baryon fields as well as the chiral invariant mass terms in the baryon sector.
Table~\ref{table:allowed operators} summarizes the Yukawa interactions allowed by the symmetry.  Here ${\mathcal W}$ denotes an interaction between the same species of baryons (e.g. $\Psi_{1l}$ and $\Psi_{1r}$), and $\mc{O}$ denotes an interaction between different species of baryons.
Their explicit forms are given in Tables~\ref{operators  W} and \ref{operators O}, respectively.
Similarly, the kinetic terms, $\mc{K}$, and the chiral invariant mass terms $\mc{Q}$ are listed in Tables~\ref{operators K} and \ref{operators Q}, respectively.
\begin{table*}[htb]
 \begin{minipage}{0.55\hsize}
\begin{center}
\caption[]{List of the allowed operators: $\mc{O}_{L, R}$.
}\label{operators O}
\begin{tabular}{c||c|c}\hline\hline
\hspace{1mm}$k$\hspace{1mm}
& $\mc{O}^{(k)}_L$
&$\mc{O}^{(k)}_R$ 
\\\hline
$1$\hspace{2mm}&$
\epsilon_{abc}\braa{\bar{\Psi}_{1r}}_\alpha^a 
\braa{M}^b_\beta\braa{\eta_{1l}}^{(c, \alpha\beta)}
$&$
\epsilon_{\alpha\beta\sigma}\braa{\bar{\Psi}_{1l}}^\alpha_a \braa{M^\dagger}_b^\beta\braa{\eta_{1r}}^{(ab, \sigma)}
$\\
$2$\hspace{2mm}&$
\tr \brac{{\bar{\Psi}_{1r}}{M}^\dagger{\chi_{1l}}}
$&$
\tr \brac{{\bar{\Psi}_{1l}}{M}{\chi_{1r}}}
$\\
$3$\hspace{2mm}&$
\epsilon_{\alpha\beta\sigma}\braa{\bar{\Psi}_{2r}}^\alpha_a 
\braa{M^\dagger}_b^\beta\braa{\eta_{2l}}^{(ab, \sigma)}
$&$
\epsilon_{abc}\braa{\bar{\Psi}_{2l}}_\alpha^a 
\braa{M}^b_\beta\braa{\eta_{2r}}^{(c, \alpha\beta)}
$\\
$4$\hspace{2mm}&$
\tr \brac{{\bar{\Psi}_{2r}}{M}{\chi_{2l}}}
$&$
\tr \brac{{\bar{\Psi}_{2l}}{M}^\dagger{\chi_{2r}}}
$\\
$5$\hspace{2mm}&$
\epsilon^{acd}\braa{\bar{\eta}_{1r}}_{(ab, \alpha)}
\braa{M^\dagger}_c^\alpha\braa{\chi_{1l}}^{b}_d
$&$
\epsilon^{\alpha\sigma\rho}\braa{\bar{\eta}_{1l}}_{(a, \alpha\beta)}
\braa{M}^a_\sigma\braa{\chi_{1r}}^{\beta}_\rho
$\\
$6$\hspace{2mm}&$
\epsilon^{\alpha\sigma\rho}\braa{\bar{\eta}_{2r}}_{(a, \alpha\beta)}
\braa{M}^a_\sigma\braa{\chi_{2l}}^{\beta}_\rho
$&$
\epsilon^{acd}\braa{\bar{\eta}_{2l}}_{(ab, \alpha)}
\braa{M^\dagger}_c^\alpha\braa{\chi_{2r}}_{d}^b
$\\\hline\hline
\end{tabular}
\end{center}
\label{operators1}
\end{minipage}
 \begin{minipage}{0.43\hsize}
\begin{center}
\caption[]{List of the allowed kinetic terms: $\mc{K}_{L,R}$.
}\label{operators K}
\begin{tabular}{c||c|c}\hline\hline
\hspace{1mm}$k$\hspace{1mm}
&$\mc{K}^{(k)}_L$&$\mc{K}^{(k)}_R$
\\\hline
$1$\hspace{2mm}&
$\tr \brac{\bar{\Psi}_{1l} i\Dsl \Psi_{1l}}$
&
$\tr \brac{\bar{\Psi}_{1r} i\Dsl \Psi_{1r}}$
\\
$2$\hspace{2mm}&
$\tr \brac{\bar{\Psi}_{2l} i\Dsl \Psi_{2l}}$
&
$\tr \brac{\bar{\Psi}_{2r} i\Dsl \Psi_{2r}}$
\\
$3$\hspace{2mm}&
$\braa{\bar{\eta}_{1l}}_{(a, \alpha\beta)} i\Dsl \braa{\eta_{1l}}^{(a, \alpha\beta)}$
&
$\braa{\bar{\eta}_{1r}}_{(ab, \alpha)} i\Dsl \braa{\eta_{1r}}^{(ab, \alpha)}$
\\
$4$\hspace{2mm}&
$\braa{\bar{\eta}_{2l}}_{(ab, \alpha)} i\Dsl \braa{\eta_{2l}}^{(ab, \alpha)}$
&
$\braa{\bar{\eta}_{2r}}_{(a, \alpha\beta)} i\Dsl \braa{\eta_{2r}}^{(a, \alpha\beta)}$
\\
$5$\hspace{2mm}&
$\tr \brac{\bar{\chi}_{1l} i\Dsl \chi_{1l}}$
&
$\tr \brac{\bar{\chi}_{1r} i\Dsl \chi_{1r}}$
\\
$6$\hspace{2mm}&
$\tr \brac{\bar{\chi}_{2l} i\Dsl \chi_{2l}}$
&
$\tr \brac{\bar{\chi}_{2r} i\Dsl \chi_{2r}}$
\\\hline\hline
\end{tabular}
\end{center}
\label{kinetic terms}
 \end{minipage}
 \begin{minipage}{0.54\hsize}
\begin{center}
\caption[]{List of the allowed operators: $\mc{W}_{L, R}$.
}\label{operators W}
\begin{tabular}{c||c|c}\hline\hline
\hspace{1mm}$k$\hspace{1mm}
&$\mc{W}^{(k)}_L$ & $\mc{W}^{(k)}_R$
\\\hline
$1$\hspace{2mm}&$
\epsilon_{abc}
\epsilon^{\alpha\beta\sigma}
\braa{\bar{\Psi}_{1r}}_\alpha^a 
\braa{M}^b_\beta
\braa{\Psi_{1l}}^c_\sigma
$&$
\epsilon^{abc}
\epsilon_{\alpha\beta\sigma}
\braa{\bar{\Psi}_{1l}}^\alpha_a 
\braa{M^\dagger}_b^\beta
\braa{\Psi_{1r}}_c^\sigma
$\\
$2$\hspace{2mm}&$
\epsilon^{abc}
\epsilon_{\alpha\beta\sigma}
\braa{\bar{\Psi}_{2r}}^\alpha_a 
\braa{M^\dagger}_b^\beta
\braa{\Psi_{2l}}_c^\sigma
$&$
\epsilon_{abc}
\epsilon^{\alpha\beta\sigma}
\braa{\bar{\Psi}_{2l}}_\alpha^a 
\braa{M}^b_\beta
\braa{\Psi_{2r}}^c_\sigma
$
\\
$3$\hspace{2mm}&$
\braa{\bar{\eta}_{1r}}_{(ab, \alpha)}
\braa{M}^a_\beta
\braa{\eta_{1l}}^{(b, \alpha\beta)}
$&$
\braa{\bar{\eta}_{1l}}_{(a, \alpha\beta)}
\braa{M^\dagger}_b^\alpha
\braa{\eta_{1r}}^{(ab, \beta)}
$
\\
$4$\hspace{2mm}&$
\braa{\bar{\eta}_{2r}}_{(a, \alpha\beta)}
\braa{M^\dagger}_b^\alpha
\braa{\eta_{2l}}^{(ab, \beta)}
$&$
\braa{\bar{\eta}_{2l}}_{(ab, \alpha)}
\braa{M}^a_\beta
\braa{\eta_{2r}}^{(b, \alpha\beta)}
$
\\\hline\hline
\end{tabular}
\end{center}
\label{operators2}
 \end{minipage}
 \begin{minipage}{0.45\hsize}
\begin{center}
\caption[]{List of the allowed mass operators: $\mc{Q}_{L,R}$.
}\label{operators Q}
\begin{tabular}{c||c|c}\hline\hline
\hspace{1mm}$k$\hspace{1mm}
&$
\mc{Q}^{(k)}_L$ & ${\mc{Q}}^{(k)}_R$
\\\hline
$1$\hspace{2mm}&$
-\tr\brac{\bar{\Psi}_{1r} \Psi_{2l}}
$&$
\tr\brac{\bar{\Psi}_{1l} \Psi_{2r}}
$\\
$2$\hspace{2mm}&$
-\braa{\bar{\eta}_{1r}}_{(ab, \alpha)} \braa{\eta_{2l}}^{(ab, \alpha)}
$&$
\braa{\bar{\eta}_{1l}}_{(a, \alpha\beta)} \braa{\eta_{2r}}^{(a, \alpha\beta)}
$\\
$3$\hspace{2mm}&$
-\tr\brac{\bar{\chi}_{1r} \chi_{2l}}
$&$
\tr\brac{\bar{\chi}_{1l} \chi_{2r}}
$
\\\hline\hline
\end{tabular}
\end{center}
\label{mass operators}
 \end{minipage}
\end{table*}

The operators transform under the parity as
\bee{
\mc{K}_{L}\lrax{\mc{P}}{}\mc{K}_{R}
\ ,~~
\mc{W}_{L}\lrax{\mc{P}}{}\mc{W}_{R}
&\ ,~~
\mc{O}_{L}\lrax{\mc{P}}{}\mc{O}_{R}
\ ,~~
\mc{Q}_{L}\lrax{\mc{P}}{}\mc{Q}_{R}
\ .
}
Thus 
one can find that
the invariant operators under the parity
are
\bee{
\mc{K}\equiv \mc{K}_{L}+\mc{K}_{R}
&\ ,~~
\mc{W}\equiv\mc{W}_{L}+\mc{W}_{R}
\ ,
\nn\\
\mc{O}\equiv\mc{O}_{L}+\mc{O}_{R}
&\ ,~~
\mc{Q}\equiv\mc{Q}_{L}+\mc{Q}_{R}
\ .
}
By requiring invariance under the parity transformation, 
we have the following Lagrangian:
\bee{
\hspace{-5mm}
\mc{L}=&
\sum_{k=1}^{6}\mc{K}^{(k)}
+\sum_{k=1}^{4} g_k \mc{W}^{(k)}
\nn\\&
+\sum_{k=1}^{6} y_k\braa{ \mc{O}^{(k)}+ {\mc{O}^{(k)}}^\dagger}
+\sum_{k=1}^{3} m_0^{(k)}\braa{\mc{Q}^{(k)}+{\mc{Q}^{(k)}}^\dagger}
\nn\\&
+\mc{L}_{\rm meson}
\label{Lagrangian}
}
where $g^k, y^k,$ and $m_0^{(k)}$ are real constants.
Note that the charge conjugation makes these parameters to be real:
\bee{
\mc{K}\lrax{C}{}\mc{K}
\ ,~~&
\mc{W}\lrax{C}{}\mc{W}
\ ,~~
\mc{O}\lrax{C}{}{\mc{O}}^\dagger
\ ,~~
\mc{Q}\lrax{C}{}{\mc{Q}}^\dagger
\ .
}

The mesonic part of the Lagrangian denoted by
$\mc{L}_{\rm meson}$
is written as
\bee{
\mc{L}_{\rm meson}=
\tr \brac{D_\mu M \cdot D^\mu M^\dagger}
-V \braa{ M }
}
where $V \braa{ M }$
is a meson potential term, which
includes the breaking of U$(1)_{\rm A}$ by the anomaly.

\subsection{Mass matrix}
We have constructed the Lagrangian by requiring the chiral U$(3)_{\rm L}$$\times$U$(3)_{\rm R}$ invariance.
To study the properties of nucleons, we decompose baryons in the chiral representations to the irreducible representation of the flavor symmetry.  
Then, we obtain the mass matrix for nucleons.

We assume that the meson potential $V \braa{ M }$ makes $M$ to have the expectation value given as
\bee{
\brae{M}=&
\braa{\mt{
v_1&0&0\\
0&v_2&0\\
0&0&v_3\\
}}
\ ,}
where we take 
$v_1=v_2=f_\pi/2$ due to the isospin symmetry.
This breaks down the chiral symmetry U$(3)_{\rm L}$$\times$U$(3)_{\rm R}$ to the flavor symmetry spontaneously.
Note that 
we do not determine the value of $v_3$ here because the value 
does not affect our results in this paper.

One can decompose each representation of the chiral symmetry to the irreducible representations of the flavor symmetry:
\bee{
&~~~~~~~~
\Psi_i  = B^{(1)}_i+ \frac{1}{\sqrt{3}}\Lambda^{(4)}_i {\bf 1}
\ ,
\nn\\&
\left\{\mt{~~
\eta^{(a, \alpha\beta)}_{1l , 2r} = \Delta^{a\alpha\beta}_{1l , 2r}
 + \frac{1}{\sqrt{6}}\braa{\epsilon^{\alpha a c}\delta^{\beta}_{k}+\epsilon^{\beta a c}\delta^{\alpha}_{k}}
\braa{B^{(2)}_{1l , 2r}}^k_c
\\
\eta^{(ab, \alpha)}_{1r , 2l} = \Delta^{ab \alpha}_{1r , 2l}
 + \frac{1}{\sqrt{6}}\braa{\epsilon^{a\alpha c}\delta^{b}_{k}+\epsilon^{b\alpha c}\delta^{a}_{k}}
\braa{B^{(2)}_{1r , 2l}}^k_c
}\right.
\ ,~~
\nn\\&~~~~~~~~
\chi_i = B^{(3)}_i
}
for $i=1l,$ $1r ,$ $2l,$ or $2r$ where $\Delta, B,$ and $\Lambda$ belong to decaplet-, octet-, and singlet-representations of the flavor symmetry, respectively.
In the following we use the Dirac spinor expression for convenience:
\bee{
B_{1} &\equiv B_{1l} +B_{1r}
\ ,~~
B_{2} \equiv B_{2l} +B_{2r}
\ ,
\nn\\
\Delta_{1} &\equiv \Delta_{1l} +\Delta_{1r}
\ ,~~
\Delta_{2} \equiv \Delta_{2l} +\Delta_{2r}
\ ,
\nn\\
\Lambda_{1} &\equiv \Lambda_{1l} +\Lambda_{1r}
\ ,~~
\Lambda_{2} \equiv \Lambda_{2l} +\Lambda_{2r}
\ .
}
Each component of the octet-baryon $B$ is assigned as
\bee{
B^a_b=&\braa{
\mt{
B_1^1&B_2^1&B_3^1\\
B_1^2&B_2^2&B_3^2\\
B_1^3&B_2^3&B_3^3
}
}
\nn\\
=&\braa{
\mt{
\frac{1}{\sqrt{2}}\Sigma^0+\frac{1}{\sqrt{6}} \Lambda & \Sigma^+& p \\
\Sigma^- & \frac{-1}{\sqrt{2}}\Sigma^0+\frac{1}{\sqrt{6}} \Lambda& n\\
\Xi^-&\Xi^0& \frac{-2}{\sqrt{6}}\Lambda
}
}
\ .
}
Similarly, for the decaplet-baryon $\Delta$ ($J=1/2$), we can set
\begin{widetext}
\bee{
&
\Delta^{111}=\Delta^{++}
\ ,~~
\Delta^{112}=\frac{1}{\sqrt{3}}\Delta^+
\ ,~~
\Delta^{122}=\frac{1}{\sqrt{3}}\Delta^0
\ ,~~
\Delta^{222}=\Delta^{-}
\ ,
\nn\\
&~~~~~~~~
\Delta^{113}=\frac{1}{\sqrt{3}}{\Sigma^{+}}^{(4)}
\ ,~~
\Delta^{123}=\frac{1}{\sqrt{6}}{\Sigma^{0}}^{(4)}
\ ,~~
\Delta^{223}=\frac{1}{\sqrt{3}}{\Sigma^{-}}^{(4)}
\ ,
\nn\\
&~~~~~~~~~~~~~~~~~~~~~
\Delta^{133}=\frac{1}{\sqrt{3}}{\Xi^{0}}^{(4)}
\ ,~~
\Delta^{233}=\frac{1}{\sqrt{3}}{\Xi^{-}}^{(4)}
\ ,
\nn\\
&~~~~~~~~~~~~~~~~~~~~~~~~~~~~~~~~~~~~
\Delta^{333}=\Omega
\ .
}

The scalar and pseudoscalar meson fields
$M$ is parametrized as
\bee{
M \equiv S + i\phi
}
where $S$ is the scalar meson nonet and the $\phi$ is the pseudoscalar meson nonet parametrized as
\bee{
\phi=
\frac{1}{\sqrt{2}}
\braa{\mt{
\frac{1}{\sqrt{2}}\pi^0+\frac{1}{\sqrt{6}}\eta_8+\frac{1}{\sqrt{3}}\eta_0
&\pi^+&K^+\\
\pi^-&-\frac{1}{\sqrt{2}}\pi^0+\frac{1}{\sqrt{6}}\eta_8+\frac{1}{\sqrt{3}}\eta_0&K^0\\
K^-&\bar{K}^0&-\frac{2}{\sqrt{6}}\eta_8+\frac{1}{\sqrt{3}}\eta_0\\
}}
\ .
}
\end{widetext}

The axial coupling of 
the F- and the D-types are 
determined by the chiral representation
from the kinetic terms in the matrix form:
\bee{
\tr \brac{\bar{B} 
\gamma_5\gamma^\mu 
\braa{D}_{6\times 6}
 \brab{A_\mu, B}}
+
\tr \brac{\bar{B} 
\gamma_5\gamma^\mu 
\braa{F}_{6\times 6}
 \brac{A_\mu, B}}
}
where
\begin{align}
\braa{F}_{6\times 6} & ={\rm diag} \braa{0,\ \frac{2}{3},\ -1,\ 0,\ -\frac{2}{3},\ 1 } \ , 
\notag\\
\braa{D}_{6\times 6} & ={\rm diag} \braa{-1,\ 1,\ 0,\ 1,\ -1,\ 0 } \ , \notag\\
{B}^T & \equiv \braa{\mt{{B}_{1}^{(1)} &{B}_{1}^{(2)}&{B}_{1}^{(3)} & {B}_{2}^{(1)}& {B}_{2}^{(2)} & {B}_{2}^{(3)}}} \ , 
\end{align}
and 
the trace is taken in the flavor space.

In the following we will not explicitly discuss the hyperon sector in which 
the contribution of the current quark mass is not negligible,
and focus the study on 
the nucleon $N(=p, n)$ and excited nucleons.

Let us derive the mass matrix.
In the following, 
for convenience, we redefine the baryon fields as
\bee{
\psi'_1 =\psi_1 
\ ,~~
\psi'_2 =\gamma_5 \psi_2 
\label{redefinition}
}
for $\psi= N^{(1)},~N^{(2)},$ or $N^{(3)}$.
Note that the definitions of $N^{(1)},~N^{(2)},$ and $N^{(3)}$ are the components of the nucleon doublets in the matrices of $B^{(1)},~B^{(2)},$ and $B^{(3)}$, respectively.
This redefinition makes all of their parities to be even: i.e. ${N'}_{1,2}^{(1)} \xrightarrow[]{\mc{P}} \gamma_0 {N'}_{1,2}^{(1)}$.

One can obtain the mass term for $N$ from the Lagrangian~\eqref{Lagrangian} as in the following form:
\bee{
-\bar{N}'M_N' N'
}
with
\begin{align}
{N'}^T &\equiv \braa{\mt{{N'}_{1}^{(1)} &{N'}_{1}^{(2)}&{N'}_{1}^{(3)} & {N'}_{2}^{(1)}& {N'}_{2}^{(2)} & {N'}_{2}^{(3)}}} \ .
\end{align}
Here the mass matrix $M_N'$ is given as
\begin{widetext}
\bee{
M'_N=&
\braa{\mt{ 
\frac{g_1}{2}f_\pi &-\frac{3y_1}{2\sqrt{6}}f_\pi&-\frac{y_2}{2}f_\pi&m_0^{(1)} &0&0
\\
-\frac{3y_1}{2\sqrt{6}}f_\pi&\frac{g_3}{4}f_\pi &-\frac{3y_5}{2\sqrt{6}}f_\pi&0&m_0^{(2)}  &0
\\ 
-\frac{y_2}{2}f_\pi&-\frac{3y_5}{2\sqrt{6}}f_\pi&0&0&0&m_0^{(3)} 
\\ 
m_0^{(1)}  &0&0& -\frac{g_2}{2}f_\pi&\frac{3y_3}{2\sqrt{6}}f_\pi&\frac{y_4}{2}f_\pi
\\ 
0&m_0^{(2)} &0&\frac{3y_3}{2\sqrt{6}}f_\pi&-\frac{g_4}{4}f_\pi &\frac{3y_6}{2\sqrt{6}}f_\pi
\\
0&0&m_0^{(3)} &\frac{y_4}{2}f_\pi&\frac{3y_6}{2\sqrt{6}}f_\pi& 0}}
\ .
}
\end{widetext}
The mass eigenstates are obtained by diagonalizing the mass matrix, which are denoted by
\bee{
{N'}^T_{\rm phys}=& \braa{\mt{{N'}_{+}^{(1)} &{N'}_{+}^{(2)}&{N'}_{+}^{(3)} & {N'}_{-}^{(1)}& {N'}_{-}^{(2)} & {N'}_{-}^{(3)}}}_{\rm phys}
\ .
}
By conducting the inverse process of Eq.~\eqref{redefinition}, 
one can easily see that
the parities of the fields with the ``$-$'' subscript are odd, i.e. $N_{\pm} \xrightarrow[\mc{P}]{} \pm \gamma_0 N_\pm$.
\begin{widetext}
In the present analysis, we identify the mass eigenstates as 
\bee{
 {N}^T_{\rm phys}=\braa{\mt{N (939)~~  N (1440)~~  N (1710)~~ 
  N (1535) ~~ N (1650) ~~  N^{\rm 6th} }}_{\rm phys}
\label{N massstate}
}
where $N^{\rm 6th}$ is an unknown baryon with odd parity.
One of its candidates is a baryon named as $N (1895) (J^P=\frac{1}{2}^-)$.
 Each of the above baryons has its eigenvalues $\lambda_{N_i}$.
\end{widetext}

\section{Extended Goldberger-Treiman relation}
\label{sec:upper bounds}
In this section we will give the extended Goldberger-Treiman relation and show that it yields an upper bound of decay widths of exited nucleons.

The interaction term is given from the Lagrangian as
\bee{
\bar{N}'C_{\pi NN}' i\gamma_5 \pi N'
}
where $\pi= \pi^a \cdot \sigma^a$ and
\bee{
C_{\pi NN}'
=
\braa{
\mt{
-\frac{g_1}{2} &-\frac{y_1}{2\sqrt{6}}&\frac{y_2}{2}&0&0&0
\\
-\frac{y_1}{2\sqrt{6}}&\frac{5g_3}{12} &-\frac{y_5}{2\sqrt{6}}&0&0&0
\\
\frac{y_2}{2} &-\frac{y_5}{2\sqrt{6}}&0&0&0&0
\\
0&0&0&-\frac{g_2}{2} &-\frac{y_3}{2\sqrt{6}}&\frac{y_4}{2}
\\
0&0&0&-\frac{y_3}{2\sqrt{6}}&\frac{5g_4}{12} &-\frac{y_6}{2\sqrt{6}}
\\
0&0&0&\frac{y_4}{2} &-\frac{y_6}{2\sqrt{6}}&0
}
}
\ .
}

Now by using the $\braa{F}_{6 \times 6},$ $\braa{D}_{6 \times 6},$ $M_N',$ and $C_{\pi NN}'$, we can find the extended Goldberger-Treiman relation:
\bee{
&C_{\pi NN}' 
=
\frac{1}{2f_\pi}
\brab{
\braa{F+D}_{6 \times 6}, M_N'
\label{extendedGT}
}
}
which is on the chiral representation basis.
We can also get the coupling constants $g_{\pi N_i N_j}'$ for the physical state :
\bee{
g_{\pi N_i N_j}'
&\equiv
u^T_i \cdot C_{\pi NN}'  \cdot u_j
\nn\\
&=
\frac{\lambda_{N_i}  + \lambda_{N_j}  }{2f_\pi}
\ 
\brac{
u^T_i \cdot
\braa{F+D}_{6 \times 6}
 \cdot
u_j 
}
}
where an eigenvector with respect to $M'_N$ is denoted by $u_i$
and its eigenvalue is $\lambda_{N_i}$.
We set $u_1$ as the eigenvector corresponding to the nucleon, $N(939)$.
One can verify that this relation is consistent with the two flavor case given in Refs.~\cite{Detar:1988kn,Jido:2001nt}.

Now the relation yields
\bee{
\sum_j\frac{f_\pi^2}
{\braa{\lambda_{N_i}  + \lambda_{N_j}  }^2}
\braa{g_{\pi N_i N_j}'}^2
=
\frac{1}{4}\brac{
u_i^T
\cdot
\braa{F+D}_{6\times6}^2
\cdot u_i 
}
\ .
}
Noting that the right hand side is bounded as
\bee{
&
\frac{1}{4}
\brac{
u^T_i\cdot
\braa{F+D}_{6\times6}^2
\cdot u_i}
\leq\ 
\frac{25}{36} \ ,
\label{limit g}
}
we obtain the following upper bound 
independently of the model parameters:
\bee{
\frac{1}{4}\braa{g_A}^2
+
\sum_{N^*}
\tilde{\Gamma}\braa{N^*}
\leq&\
\frac{25}{36}
\label{upper bound}
}
where $g_A$ is the axial coupling of the nucleon $N(939)$ given by
\bee{
g_A = u_1^T\cdot \left( F + D \right)_{{6\times 6}}\cdot u_1
\ ,
}
and $\tilde{\Gamma}(N^*)$ is defined as
\footnote
{The partial decay width of $N^* \rightarrow N + \pi$ is given by
\bee{
\Gamma (N^* \rightarrow N(939)+ \pi)
=& 
\frac{3g_{\pi N^*N}^2 }{4\pi }\frac{\brad{\vec{p}}}{m_{N^*}}\brac{ \sqrt{\brad{\vec{p}}^2+m_N^2} \mp m_N}
\nn
}
where the minus sign in the bracket is chosen if the parity of $N^*$ is positive. 
$\brad{\vec{p}}$ is the momentum of the nucleon, which satisfies
$m_{N^*}=\sqrt{\brad{\vec{p}}^2+m_N^2} + \sqrt{\brad{\vec{p}}^2+m_\pi^2}$.
}
\bee{
\tilde{\Gamma} \braa{N^*}\equiv&
\frac{4\pi}{3}
\frac{m_{N^*}}{\brad{\vec{p}}}\brac{ \sqrt{\brad{\vec{p}}^2+m_{N}^2} \mp m_{N}}^{-1}
\frac{f_\pi^2}{\braa{m_{N^*} \pm  m_{N}  }^2}
\nn\\
&\times 
\Gamma (N^* \rightarrow N + \pi)
\ .
\label{tilde Gamma}
}

In the following, 
to discuss how the upper bound works, we set $\tilde{\Gamma}\braa{N^{\rm 6th}}=0$.
By using the experimental values in Table~\ref{physical values}, the pion mass $m_\pi=137$~MeV, and the pion decay constant $f_\pi=92.4$~MeV together with 
$g_A = 1.268 \pm 0.006$~\cite{Yamanishi:2007zza}, one can estimate
\bee{
\frac{1}{4} g_A^2 &= 0.402 \pm 0.004
\ ,
\nn\\
\tilde{\Gamma}^{\rm exp}\braa{N \braa{1440}} &= 0.063 \ \pm\ 0.022
\ ,
\nn\\
\tilde{\Gamma}^{\rm exp}\braa{N \braa{1535}} &= 0.012 \ \pm\  0.003
\ ,
\nn\\
\tilde{\Gamma}^{\rm exp}\braa{N \braa{1650}} &= 0.010 \ \pm\ 0.004
\ ,
\nn\\
\tilde{\Gamma}^{\rm exp}\braa{N \braa{1710}} &= 0.0011 \ \pm\ 0.0018
\ ,
\label{tilde Gamma exp}
}
which yields
\begin{equation}
\frac{1}{4}\braa{g_A}^2
+
\sum_{N^*}
\tilde{\Gamma}\braa{N^*} = 0.49 \ \pm\ 0.02 \ .
\end{equation}
Since the value of $25/36$ approximately equals to $0.69$, 
the central value from the experiments 
is allowed.

\begin{figure}[ht]
 \begin{center}
  \includegraphics[width=90mm]{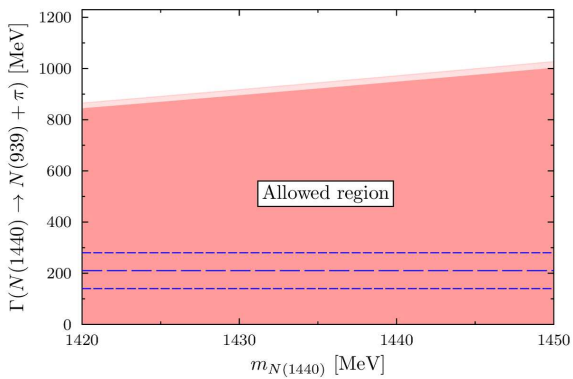}
 \end{center}
 \caption[]{
Allowed region for the $N (1440) \rightarrow N(939) + \pi$ decay width and the mass of $N(1440)$ obtained from Eq.~\eqref{constraint1}.
The experimental values of $m_{N(1440)}$ and $\Gamma \braa{N^* \rightarrow N + \pi }$ are $1420$ -- $1450$\,MeV and $210 \pm 70$\,MeV, respectively.
}
 \label{fig1}
\end{figure}

If we use the values of Eq.~\eqref{tilde Gamma exp} except $\tilde{\Gamma}^{\rm exp}\braa{N (1440)}$,
Eq.~(\ref{upper bound}) implies that
\bee{
\tilde{\Gamma}\braa{N (1440)}
\leq&\
0.269 \pm 0.007
\ 
\label{constraint1}
}
must be satisfied.
Since $\tilde{\Gamma}\braa{N \braa{1440}}$ is a function of $m_{N(1440)}$ and ${\Gamma}\braa{N \braa{1440}\rightarrow N + \pi }$
as given in Eq.~(\ref{tilde Gamma}), 
this inequality gives an allowed region on the mass of ${N(1440)}$ versus ${\Gamma}\braa{N \braa{1440}\rightarrow N + \pi }$ plane as shown in FIG.~\ref{fig1}.
This shows that 
the experimental value $210 \pm 70$ MeV~\cite{Agashe:2014kda} is 
consistent with the upper bound of the decay width of $N(1440) \rightarrow N(939) + \pi$ independently of the detail of the model parameters. 
\begin{table*}[hbt]
\caption[]{Experimental values of Masses and partial decay widths of baryons~\cite{Agashe:2014kda}.
The error of $m_{N(939)}$ expresses the mass splitting between the proton and the neutron.
}
\begin{center}
\begin{tabular}{c|ccc}\hline\hline
Baryon&Mass (MeV)&Partial width (MeV)
$\Gamma_{N^*\rightarrow N\pi}$
\\\hline
\hspace{2mm}
$N(939)$
\hspace{2mm}
&
\hspace{2mm}
$939.0 \pm1.3$
\hspace{2mm}
&---&
\\
\hspace{2mm}
$N(1440)$
\hspace{2mm}
&
\hspace{2mm}
$1420$ -- $1450$
$(\approx 1430)$
\hspace{2mm}
&
\hspace{2mm}
~~$\sim$~~
$210\pm 70$
\hspace{2mm}
\\
\hspace{2mm}
$N(1535)$
\hspace{2mm}
&
\hspace{2mm}
$1525$ -- $1545$
$(\approx 1535)$
\hspace{2mm}
&
\hspace{2mm}
~~$\sim$~~
\ $70\pm 19$
\hspace{2mm}
\\
\hspace{2mm}
$N(1650)$
\hspace{2mm}
&
\hspace{2mm}
$1645$ -- $1670$
$(\approx 1655)$
\hspace{2mm}
&
\hspace{2mm}
~~$\sim$~~
$100\pm 35$
\hspace{2mm}
\\
\hspace{2mm}
$N(1710)$
\hspace{2mm}
&
\hspace{2mm}
$1680$ -- $1740$
$(\approx 1710)$
\hspace{2mm}
&
\hspace{2mm}
~~$\sim$~~
\ $13$$^{+20}_{-13}$
\hspace{2mm}
\\\hline\hline
\end{tabular}
\end{center}
\label{physical values}
\end{table*}

\section{Numerical Analysis}
\label{sec:Numerical fitting}

In the previous section, we obtained the upper bound of the decay width of $N(1440)$ 
independently of the model parameters. 
In this section, we make a numerical analysis for fitting the model parameters and make
several physical quantities in addition to $\Gamma(N(1440))$.

There are thirteen parameters in the model:
\bee{
&
m_0^{(1)}
\ ,~~
m_0^{(2)}
\ ,~~
m_0^{(3)}
\ ,~~
\nn\\&
g_1
\ ,~~
g_2
\ ,~~
g_3
\ ,~~
g_4
\ ,
\nn\\
&
y_1
\ ,~~
y_2
\ ,~~
y_3
\ ,~~
y_4
\ ,~~
y_5
\ ,~~
y_6
\ .
}
To determine the parameters, we use ten inputs which are $F^{\rm exp}= 0.475 \pm 0.004,$ $D^{\rm exp}=0.793 \pm 0.005,$ and the experimental values of the masses and decay widths of baryons listed in Table~\ref{physical values} except $\Gamma^{\rm exp} \braa{N(1440 )}$.
Note that we cannot use the decay width $\Gamma^{\rm exp} \braa{N(1440 )}$ as an independent input because of the existence of the extended Goldberger-Treiman relation in Eq.~\eqref{extendedGT}.
In the following, 
we identify the exited nucleon $N^{\rm 6th} (J^P=\frac{1}{2}^-)$ as $N(1895)$. 

\begin{figure}[ht]
 \begin{center}
  \includegraphics[width=90mm]{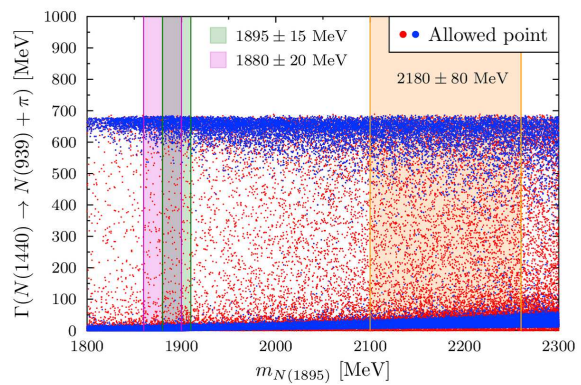}
 \end{center}
 \caption[]{Allowed values of $\Gamma (N(1440) \rightarrow N(939) + \pi)$ shown by red and blue points.
Purple, green, and orange areas are for $m_{N(1895)}= 1880\pm20$\,MeV, 
$1895\pm15$\,MeV, and $2180\pm80$\,MeV, respectively. 
}
 \label{fig7}
\end{figure}

There are still three parameters left undetermined even when we use the above ten inputs.
Nevertheless, there exists an upper bound for the $N(1440) \to N(939) +\pi$ decay width due to the extended Goldberger-Treiman relation as discussed in the previous section.
In Fig.~\ref{fig7}, 
we show the allowed values of $\Gamma(N(1440)\to N(939)+\pi)$ depending on the mass of $N(1895)$.~\footnote{%
In the PDG\cite{Agashe:2014kda}, 
three experimental values are listed: $1895\pm15$, $2180\pm80$, $1880\pm20$.
}
Here each red and blue point is plotted by arbitrary choosing three undetermined parameters.
This shows that the upper bound is about $700$\,MeV almost independently of the mass of $N(1895)$.

\begin{figure}[ht]
 \begin{center}
 \includegraphics[width=80mm]{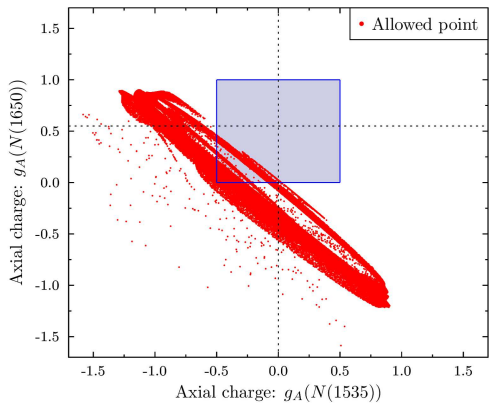}
 \end{center}
 \caption[]{Allowed values of 
$g_A (N(1535))$ and $g_A (N(1650))$ shown by red points.
Blue area shows the region restricted by
$\brad{g_A (N(1535))} \leq 0.5 $ and $0.0 \leq  {g_A (N(1650))} \leq  1.0 $.
}
 \label{fig6}
\end{figure}

Each arbitrary choice of three undetermined parameters also give the predictions for the axial couplings of the excited nucleons.
Here, restricting the parameter region from $\Gamma (N(1440) \rightarrow N(939) + \pi) = 210 \pm 70$\,MeV as an input, we show allowed values of 
$g_A (N(1535))$ and $g_A (N(1650))$ in Fig.~\ref{fig6}.
This shows that there exist upper bounds for both axial couplings:
$g_A(N(1535)) \lesssim 0.9$, $g_A(N(1650))\lesssim0.9$ and $g_A(N(1535)) +g_A(N(1650)) \lesssim 0.0$, which implies that one of $g_A(N(1535))$ and $g_A(N(1650))$ must be negative. We think that these are the consequences of some relations similar to the extended Goldberger-Treiman relation in Eq.~\eqref{extendedGT}.
We would like to stress that the above predictions shown in Fig.~\ref{fig6} will be tested by future experiments.

We make further restriction of the parameters by  referring to the lattice QCD analysis on the axial charges of $N(1535)$ and $N(1650)$ in Ref.~\cite{Takahashi:2008fy}, which shows that $g_A(N(1535)) = {\mathcal O}(0.1)$ and $g_A(N(1650)) \sim 0.55$.
In the following we use the parameter sets 
satisfying $\brad{g_A (N(1535))} \leq 0.5 $ and $0.0 \leq  {g_A (N(1650))} \leq  1.0 $ which are illustrated by the blue area in Fig.~\ref{fig6}.

Furthermore, 
we fix $m_{N(1895)}=1895$\,MeV which is the result of the newest experiment for the exited nucleon $N(1895)$ listed in PDG\,\cite{Agashe:2014kda}. 
In the following, by using it, we discuss the axial charges for each excited nucleon 
as well as the mixing structure and the chiral invariant mass of the nucleon $N(939)$.

\begin{figure}[ht]
 \begin{center}
  \includegraphics[width=90mm]{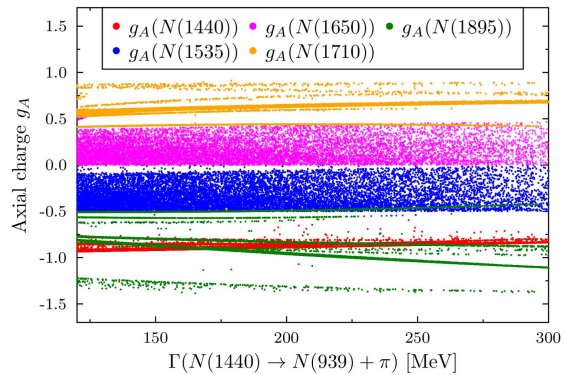}
 \end{center}
 \caption[]{
Dependence of the axial charges of exited nucleons on $\Gamma (N(1440) \to N(939) +\pi) $.
}
 \label{fig2}
\end{figure}

In Fig.~\ref{fig2}, 
we plot the dependence of the values of 
the axial charges of excited nucleons on the value of the decay width of $N(1440)\to N(939) +\pi$ for $\Gamma(N(1440))= 120$ -- $300$\,MeV.
One can see that the values of the axial charges are rather stable against the value of the decay width.
As we stated above, this reflects existence of the extended Goldberger-Treiman relations for these quantities.

We also study the mixing structure of the ground state nucleon $N(939)$.
\begin{figure}[htb]
 \begin{center}
  \includegraphics[width=90mm]{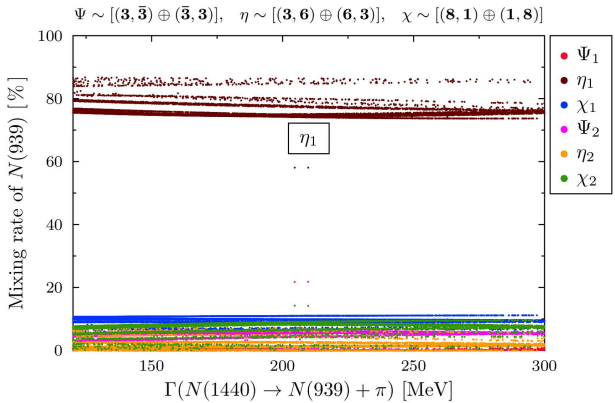}
 \end{center}
 \caption[]{
Mixing structure of the ground state nucleon $N(939)$.
The vertical axis shows the percentage of the component fields in $N(939)$.
$\Psi$, $\eta$, and $\chi$ belong to 
$\brac{({\bf 3} , \bar{{\bf 3}})\oplus (\bar{{\bf 3}} , {\bf 3})}$,
$\brac{({\bf 3} , {{\bf 6}})\oplus ({{\bf 6}} , {\bf 3})}$, and 
$\brac{({\bf 8} , {{\bf 1}})\oplus ({{\bf 1}} , {\bf 8})}$, respectively.
}
 \label{fig4}
\end{figure}
In Fig.~\ref{fig4}, we plot the percentages of the component fields included in $N(939)$ obtained from the orthogonal matrix which diagonalize the mass matrix. 
This shows that $\eta_1$ belonging to 
$\brac{({\bf 3} , {\bf 6}) \oplus ({\bf 6}, {\bf 3})}$ representation
is a dominant component ($\sim 80$\,\%) in the $N(939)$ state.
Its reason is that only the axial charge of $\eta_1$ is larger than $1$.

\begin{figure}[htb]
 \begin{center}
  \includegraphics[width=90mm]{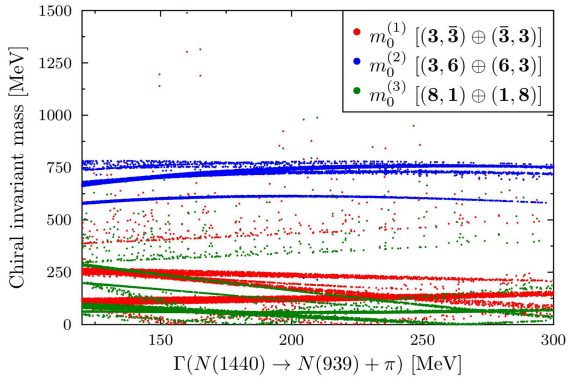}
 \end{center}
 \caption[]{
Chiral invariant masses: $m_0^{1}$, $m_0^{2}$, and $m_0^{3}$.
}
 \label{fig8}
\end{figure}

Finally, we show the values of the chiral invariant masses 
in Fig.~\ref{fig8}.
Here, 
$m_0^{(1)}$, $m_0^{(2)}$, and $m_0^{(3)}$ are the chiral invariant masses of 
$\brac{({\bf 3} , {\bf \bar{3}}) \oplus ({\bf \bar{3}}, {\bf 3})}$, 
$\brac{({\bf 3} , {\bf 6}) \oplus ({\bf 6}, {\bf 3})}$, and $\brac{({\bf 8} , {\bf 1}) \oplus ({\bf 1}, {\bf 8})}$ representations, respectively.
Figure~\ref{fig8} shows that all the values are rather stable against the change of the value of $\Gamma(N(1440) \to N(939)+\pi)$, and 
$m_0^{(2)} \sim 550$ --  $800$\,MeV. 
Since the main components of the nucleon, $N(939)$ are $\brac{({\bf 3} , {\bf 6}) \oplus ({\bf 6}, {\bf 3})}$,
as we show above,
our results imply that 
the chiral invariant mass of the nucleon $N(939)$ is roughly $500$ -- $800$\,MeV, which is consistent with the result in Ref.~\cite{Motohiro:2015taa}.

\section{Summary and Discussions}
\label{sec:summary}

We introduced three kinds of baryons belonging to $\brac{({\bf 3} , \bar{{\bf 3}})\oplus (\bar{{\bf 3}} , {\bf 3})}$, $\brac{({\bf 3} , {\bf 6}) \oplus ({\bf 6}, {\bf 3})}$, and $\brac{({\bf 8} , {\bf 1}) \oplus ({\bf 1} , {\bf 8})}$ representations of the chiral $\mbox{U}(3)_{\rm L}\times \mbox{U}(3)_{\rm R}$ group together with their parity partners.
We constructed a model Lagrangian based on the chiral U$(3)_{\rm L}$$\times$U$(3)_{\rm R}$ symmetry as well as the 
invariance under the parity and the charge conjugation.
We derived an extended Goldberger-Treiman relation, from which we obtained an upper bound for decay width of the exited nucleon $N(1440)$ as shown in 
Fig.~\ref{fig1}.

We performed a numerical analysis by assuming 
$N(939)$, $N(1440)$, $N(1535)$, $N(1650)$, $N(1710)$, and $N(1895)$ make a parity doubling structure.
Our results show that
the upper bound of $\Gamma (N(1440))$ is about $700$\,MeV, which is consistent with its experimental value $210\pm 70$\,MeV.
Furthermore, we showed that there exist upper bounds for the axial couplings of $N(1535)$ and $N(1650)$:
$g_A(N(1535)) \lesssim 0.9$, $g_A(N(1650))\lesssim0.9$ and $g_A(N(1535)) +g_A(N(1650)) \lesssim 0.0$, which implies that one of $g_A(N(1535))$ and $g_A(N(1650))$ must be negative.
These constraints can be tested by future experiments as well as lattice QCD analyses.
The violation of  our constraints
means that the present parity doubling structure may not work, and that
we may have to modify the structure of the parity doublets.
There exist two candidates of the modification:
(i) No parity partner of $N(939)$ exists.
Although the excited nucleons have parity partners, the lowest-lying nucleon does not.
(ii) Some nucleons considered in the present analysis are molecular or penta-quark states.
If so, we have to take account of the representations of molecular or penta-quark type
expressed by the direct product of five quarks.
Such representations include
the octet-baryons belonging to
\bee{
&
(\bar{\bf 3}, {\bf 3})_{-1}
~\ ,~~
({\bf 6}, {\bf 3})_{-1} 
~\ ,~~
 (\bar{{\bf 15}}, {\bf 3})_{-1}  
~\ ,~~
\nn\\&
 ({\bf 3} , \bar{{\bf 3}})_{-5}
~\ ,~~
(\bar{{\bf 6}},\bar{{\bf 3}} )_{-5}
~\ ,~~
({{\bf 15}}, \bar{{\bf 3}})_{-5}  
~\ ,~~
\nn\\&
(\bar{{\bf 6}},\bar{{\bf 3}})_{+1}
~\ ,~
({\bf 15},\bar{{\bf 3}})_{+1}  
\, \ ,~~
 ({\bf 15}, {\bf 6})_{+1} 
~\ ,~~
\nn\\&
 ({\bf 8},{\bf 1})_{-3} 
~\ ,~~
({\bf 1} ,{\bf 8})_{-3} 
~\ ,~~~
({\bf 8} ,{\bf 8})_{-3}
\, ~\ ,~~
 ({\bf 10} ,{\bf 8})_{-3}
}
and their parity partners.

We make further restriction of the parameters by using 
$m_{N(1895)}=1895$\,MeV as well as $\brad{g_A (N(1535))} \leq 0.5 $ and $0.0 \leq  {g_A (N(1650))} \leq  1.0 $ as the constraints, to study the mixing  structure of the ground state nucleon $N(939)$.
Our results show that
 $N(939)$ is made of about 80\% of $\brac{({\bf 3} , {\bf 6}) \oplus ({\bf 6}, {\bf 3})}$ 
component, and that
the chiral invariant mass of $N(939)$ is roughly $500$ -- $800$\,MeV.
This implies that the origin of the mass of $N(939)$ is almost the chiral invariant mass, which seems consistent with the lattice QCD results~\cite{Glozman:2012fj,Aarts:2015mma}.

In addition to the Lagrangian given in Eq.~\eqref{Lagrangian}, we can add the explicit chiral symmetry braking terms due to the current quark masses.
The mass splitting among the baryons included in an octet is estimated 
as about $200$\,MeV due to the flavor symmetry breaking mainly caused by the strange quark mass.
Then, one can expect that the coefficients for the explicit symmetry breaking terms are not large and 
the modification of the present results for non-strange sector 
will be only slightly.
On the other hand, 
the effect of the flavor symmetry braking becomes important for hyperons.
We leave the study of hyperons for future works. 

\section*{ACKNOWLEDGEMENTS}
This work was supported in part by the JSPS Grant-in-Aid for Scientific Research (C) No.~24540266.

\end{document}